\documentclass[cameraready]{Interspeech}

\title{Cross-Lingual Speaker Verification with Self-Supervised Pre-Trained Models}

\author[equalcontribution]{Jinghan}{Peng}
\author[equalcontribution]{Yu}{Zheng}
\author[]{Weiqiang}{Wang}
\author[correspondingauthor]{Jian}{Liu}

\address{
    Machine Intelligence, Ant Group, Shanghai, China
}

\email{\{pengjinghan.pjh,zy466683,weiqiang.wwq,rex.lj\}@antgroup.com}


\usepackage{comment}

\usepackage{multirow} 
\usepackage{pifont} 
\usepackage{cite} 

\begin{document}

\maketitle

\begin{abstract}
    Speaker verification (SV) performance degrades under language mismatch due to the entanglement of speaker identity with language-specific acoustic cues.
    To address this problem, we leverage large-scale self-supervised pre-trained models (PTMs) to learn language-agnostic speaker representations. 
    We utilize PTMs as robust front-end feature extractors, capitalizing on their rich acoustic and linguistic knowledge acquired from vast, diverse audio data. These generalized features are then used to train a downstream speaker embedding network, effectively disentangling speaker identity from language-specific characteristics. 
    We validate our approach on the TidyVoice2026 benchmark, which benchmarks SV under language mismatch. Our proposed system (team~T02) achieves equal error rates (EERs) of 2.21\% on tv26\_eval-A and 2.99\% on tv26\_eval-U.

\end{abstract}

\section{Introduction}

Speaker verification (SV), the task of confirming a claimed identity based on a speaker’s voice, has witnessed substantial advancements over the past decade~\cite{snyder2018x,yu2020densely,wang2023cam++,desplanques2020ecapa,lin2024voxblink2,yakovlev24_interspeech}.
However, the real-world deployment of SV systems in our globalized world hinges on their ability to handle a critical bottleneck: language mismatch~\cite{zheng20233d,farhadipour2026tidyvoice}. 
This prevalent condition in multilingual scenarios, where a speaker's voice characteristics are confounded by the acoustic properties of different languages, presents a formidable challenge. 
Consequently, recognition accuracy often degrades significantly when enrollment and verification occur in different languages. 
The TidyVoice2026 benchmark~\cite{farhadipour2026tidyvoice} is established to tackle this very issue of speaker verification under language mismatch.
The freely accessible TidyVoice2026 benchmark tackles this issue by providing a multilingual dataset, driving reproducible research toward effective cross-lingual solutions.

The central challenge in cross-lingual SV is to extract a language-agnostic speaker representation that encapsulates the unique physiological and behavioral traits of a speaker while remaining invariant to the linguistic content being spoken. 
Recent breakthroughs in self-supervised learning (SSL) for audio offer a promising pathway. 
Pre-trained models (PTMs) like Wav2Vec 2.0~\cite{baevski2020wav2vec}, XLS-R~\cite{babu2021xls}, MMS~\cite{pratap2024scaling}, HuBERT~\cite{hsu2021hubert}, WavLM~\cite{chen2022wavlm}, CLAP~\cite{elizalde2023clap}, w2v-BERT~\cite{chung2021w2v}, w2v-BERT~2.0~\cite{barrault2023seamlessm4t},  pre-trained on millions of hours of unlabeled, multilingual speech, learn rich, hierarchical representations that capture general acoustic, phonetic, and even semantic information.
Our core hypothesis is that these SSL representations provide a superior foundation for building robust cross-lingual SV systems~\cite{chen2022wavlm}. 

Rather than training an SV model from scratch on potentially limited and language-specific data, a promising alternative is to leverage the generalized knowledge embedded within large-scale pre-trained models~\cite{chen2022large,zhao2024whisper,li2025enhancing}.

Chen et al.~\cite{chen2022large} employed PTMs like Wav2Vec 2.0~\cite{baevski2020wav2vec}, HuBERT~\cite{hsu2021hubert}, and UniSpeech-SAT~\cite{chen2022unispeech}, using learnable weights to average the representations from all hidden layers, which were then fed as input features into an ECAPA-TDNN~\cite{desplanques2020ecapa} to build the speaker model. Zhao et al.~\cite{zhao2024whisper} utilized the encoder of Whisper Large-v2~\cite{radford2023robust} as the PTM and proposed Partial Multi-scale Feature Aggregation to concatenate output features from different layer modules. Similarly, Li et al.~\cite{li2025enhancing} adopted the w2v-BERT 2.0 as the PTM and used Multi-scale Feature Aggregation to concatenate features from all layers before feeding them into the downstream module.
These studies have validated the effectiveness of PTM-based SV on the VoxCeleb benchmark~\cite{Nagrani_2017,chung2018voxceleb2} and CN-Celeb benchmark~\cite{fan2020cn}. 

In this work, we employ w2v-BERT 2.0 as a front-end feature extractor, apply multi-scale feature aggregation over the layer-wise outputs, feed the aggregated representations into a pooling layer to obtain utterance-level features, and finally project them into a fixed-dimensional speaker embedding.

To ensure stable training of this self-supervised learning PTM-based speaker verification model, we effectively leverage a three-stage training paradigm.
Additionally, we experiment with the lightweight ReDimNet~\cite{yakovlev24_interspeech} as an auxiliary system for comparative analysis and subsequent score-level fusion.
Our systems achieve an Equal Error Rate (EER) of 2.21\% on the tv26\_eval-A set and 2.99\% on the tv26\_eval-U set.

Here are the key contributions of our work:

\begin{itemize}
    \item We present a powerful SV pipeline for cross-lingual verification, using a pre-trained w2v-BERT 2.0 front-end for language-agnostic speaker embeddings and a lightweight back-end for efficient speaker embedding extraction.
    
    \item  We demonstrate strong performance on the TidyVoice2026 benchmark, achieving an EER of 2.21\% on tv26\_eval-A and 2.99\% on tv26\_eval-U, validating the effectiveness of our approach under language mismatch.
    
    \item We offer a detailed analysis of key design elements, including pre-trained model selection, feature aggregation, and training strategies, providing practical guidance on model selection and training for real-world applications.

\end{itemize}

\section{Datasets}
\subsection{Training dataset}
\label{sec:train_set}

For training speaker verification models, we adopt a large-scale, multilingual collection of corpora to enhance robustness under language mismatch. 
Specifically, we utilize the \textbf{TidyVoiceX\_Train}~\cite{farhadipour2026tidyvoice,farhadipour2026tidyvoicecuratedmultilingualdataset} dataset, which contains 3,666 speakers across 40 languages and serves as a challenge-specific in-domain resource.
We incorporate the \textbf{VoxCeleb2}~\cite{chung2018voxceleb2} development set (5,994 speakers) along with the \textbf{VoxBlink2}~\cite{lin2024voxblink2} dataset (111,284 speakers), both collected from open-source online media.
We include the \textbf{CN-Celeb~1\&2}~\cite{fan2020cn} training datasets, covering 2,793 Chinese speakers collected from open-source online media, similar in nature to VoxCeleb2 but focused on Mandarin content across diverse environments. 
We utilize the \textbf{WenetSpeech}~\cite{zhang2022wenetspeech} corpus and, following the data filtering procedure of Zhang et al.~\cite{zhang2025quantifying}, derive a cleaned subset containing 35,285 speakers.
We further incorporate the \textbf{3D-Speaker}~\cite{zheng20233d} training dataset, which provides 10,000 Chinese speakers with rich variations in Chinese dialects.

We employ on-the-fly data augmentation to improve model robustness to acoustic variability. During training, a random set of augmentations is applied to each utterance on a per-utterance basis from the following four types:
\begin{itemize}
    \item  \textbf{Reverberation}: artificial reverberation is applied by convolving the signal with simulated room impulse responses (RIRs), using RIRs from the AIR dataset~\cite{ko2017study};
    \item  \textbf{Music}: a music file without vocals is randomly selected from MUSAN~\cite{snyder2015musan}, trimmed or looped to match the utterance duration, and mixed with the original signal at a signal-to-noise ratio (SNR) between 5–15 dB; 
    \item  \textbf{Noise}: generic noise segments from MUSAN are added at one-second intervals with an SNR ranging from 0–15 dB; 
    \item  \textbf{Babble}: overlapping speech from MUSAN is mixed with the original signal at an SNR of 10–30 dB to simulate crowd or conversational environments. 
\end{itemize}

During training, each utterance has a probability of 0.3 to be reverberated, 0.1 to be mixed with music, 0.1 to be corrupted with noise, and 0.1 to be mixed with babble speech.

\subsection{Development dataset}

For system development and validation, we use the \textbf{TidyVoiceX\_Dev}~\cite{farhadipour2026tidyvoice,farhadipour2026tidyvoicecuratedmultilingualdataset} dataset, which contains 808 speakers spanning 40 languages. 
The development set provides a large-scale trial list comprising 12 million verification pairs, constructed to cover different language- and speaker-mismatch scenarios.
We rely exclusively on this development set for model selection, threshold adjustment, and score calibration, ensuring that our system is optimized under the same cross-lingual conditions as the evaluation set.

\subsection{Evaluation dataset}

The evaluation is performed on the \textbf{TidyVoiceX2}~\cite{farhadipour2026tidyvoice} dataset, featuring two trial lists: \textbf{tv26\_eval-A} for partial language mismatch (enrollment: seen languages; test: unseen languages), and the more challenging \textbf{tv26\_eval-U} for full generalization, where both enrollment and test use a separate set of 38 languages unseen in training and development data.

\section{Methodology}

\subsection{Model architecture}

\begin{figure*}[htbp]
  \centering
  \includegraphics[width=\linewidth]{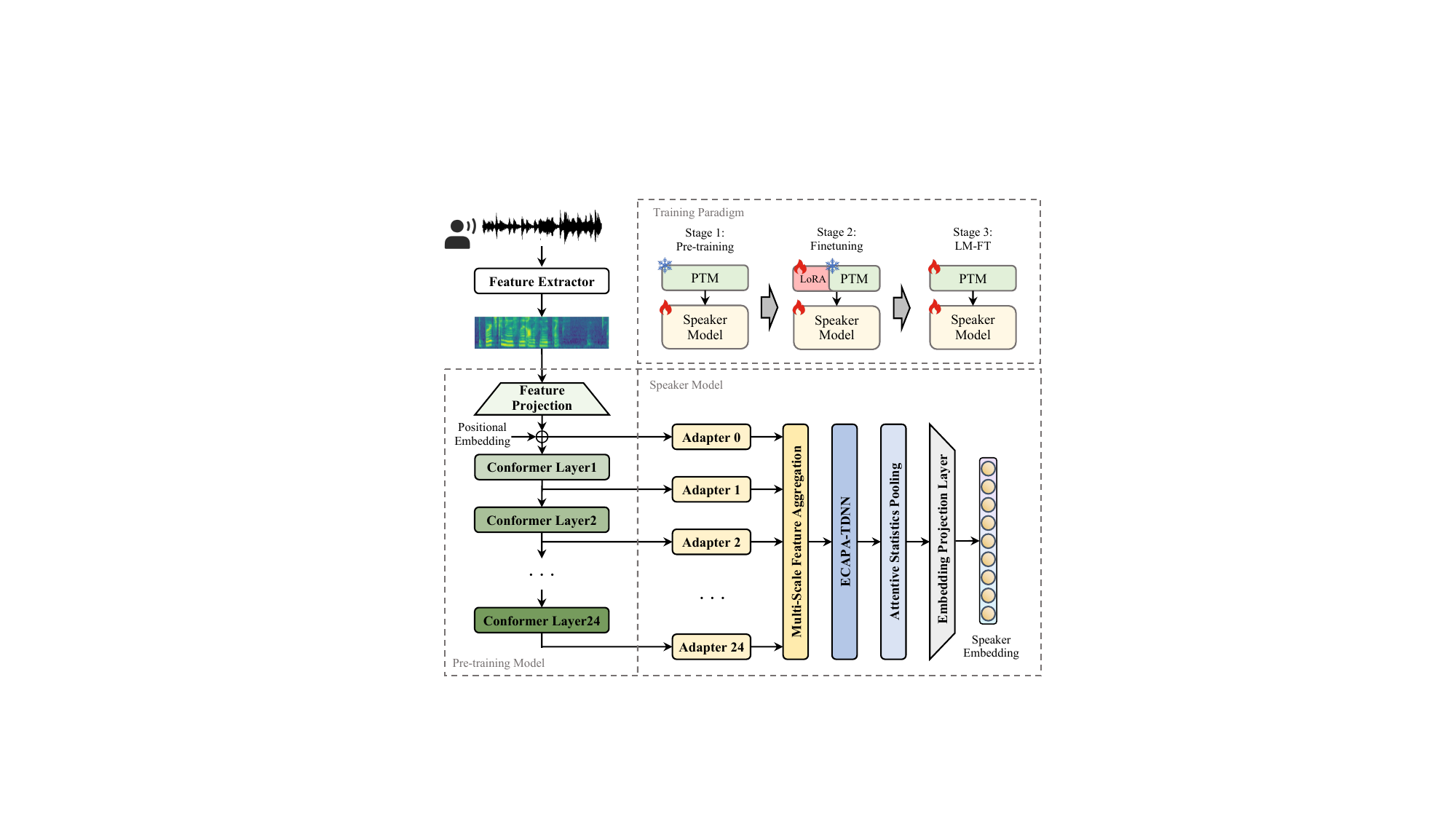}
  \caption{Model architecture and training paradigm of the speaker verification system based on self-supervised pre-trained models.}
  \label{fig:ssl_sv}
\end{figure*}

\subsubsection{PTM-based speaker verification systems}
\label{sec:ssl_ptm_sv}
The overall architecture of our primary PTM-based speaker verification system is illustrated in Figure~\ref{fig:ssl_sv}(a). 
The pipeline begins with an input audio signal, from which acoustic features such as log-Mel filterbanks are extracted. 
These features are then fed into a pre-trained SSL model (e.g., w2v-BERT 2.0~\cite{barrault2023seamlessm4t}). 
The PTM typically consists of a stack of transformer or conformer layers~\cite{barrault2023seamlessm4t,radford2023robust,pratap2024scaling,babu2021xls}. 
We extract the hidden representations from all layers of the SSL model.
To align and project the heterogeneous features from different SSL layers, each layer's output is passed through a dedicated \textit{Adapter}. Each Adapter is implemented as a two-layer Multilayer Perceptron (MLP), which projects the features from their original dimension to a unified dimension of 256.

A core component of our framework is the \textit{Multi-scale Feature Aggregation} module, which fuses the adapted features from all SSL layers. We investigate six distinct fusion mechanisms: 
\begin{enumerate}
    \item \textbf{Mean Fusion (MF)}: The unified representation is obtained by averaging the features across all layers. 
    \item \textbf{Layer-wise Weighted Fusion (LWF)}: A learnable weight is assigned to each layer, and a weighted sum of the layer features is computed. 
    \item \textbf{Channel-wise Weighted Fusion (CWF)}: Each channel in each layer is assigned a learnable weight for a fine-grained weighted fusion. 
    \item \textbf{Channel Concatenation (CC)}: Features from all layers are concatenated along the channel dimension. 
    \item \textbf{Temporal Concatenation (TC)}: Features from all layers are concatenated along the temporal dimension. 
    \item \textbf{Hierarchical Cross-Attention Fusion (HCAF)}: This method leverages a two-stage attention mechanism. First, it performs cross-attention where the representation from the last SSL layer serves as the queries, interacting with keys and values from all preceding layers to integrate multi-scale information. This is followed by a self-attention layer on the aggregated output to refine the contextual relationships and produce the final fused representation.
\end{enumerate}

The resulting frame-level features from the aggregation module are then processed by an \textit{Attentive Statistics Pooling}~\cite{okabe2018attentive} layer to generate a fixed-dimensional utterance-level representation. This representation is subsequently passed through a \textit{Embedding Projection Layer} to produce the final 256-dimensional speaker embedding.
For training, we employ the \textit{Additive Angular Margin (AAM) Softmax Loss}~\cite{deng2019arcface} to optimize the network parameters.

\subsubsection{ReDimNet systems}
For comparative analysis and system fusion, we also implement \textit{ReDimNet} models~\cite{yakovlev24_interspeech}. Specifically, we experiment with two variants: \textit{ReDimNet-B5} and \textit{ReDimNet-B6}. 
Both models produce 192-dimensional speaker embeddings and are trained using the \textit{SphereFace2-C (SF2-C)}~\cite{han2023exploring} loss function.

\subsection{Training paradigm and details}
\label{sec:train_paradigm}

In this work, we adopt different training paradigms for the PTM-based speaker verification models and RedimNet models, respectively. 
For the PTM-based speaker verification model, we employ a three-stage training strategy, as illustrated in Figure~\ref{fig:ssl_sv}(b). 
The model is initialized with the pre-trained weights from the SSL foundation model, while the remaining downstream parameters are randomly initialized. 
In the first stage, the weights of the SSL backbone are frozen, and only the downstream modules are trained using the full dataset described in Section~\ref{sec:train_set}. The training lasts for 60 epochs with 2-second segments, a batch size of 16,384, and an AAM-softmax margin penalty that gradually increases from 0 to 0.2 between epochs 20 and 30. 
The learning rate warms up for 6 epochs, starting from 0.1 and then decaying to 5e-5. 
The checkpoint with the lowest EER on the TidyVoiceX\_Dev set is selected for the next stage. 
In the second stage, the PTM backbone is unfrozen and trained for 20 epochs using the full dataset.
The audio segment length remains 2 seconds in this stage, with a fixed margin of 0.2 and a batch size of 16,384.
The learning rate warms up for 2 epochs and decays from 1e-4 to 1e-6. 
The third stage performs large-margin fine-tuning (LM-FT)~\cite{liu2019largemarginsoftmaxloss} solely on the TidyVoiceX\_Train set. Models are trained for 6 epochs with 6-second segments, a margin of 0.5, and a batch size of 512.
The learning rate is linearly warmed up to 1e-4 during the first epoch and then decayed to 1e-6 for the remaining epochs.

For the ReDimNet models, we follow the original two-stage training protocol~\cite{yakovlev24_interspeech}.
In the pre-training stage, we use the full training set with 2-second segments for 60 epochs, with the margin increasing from 0 to 0.2 between epochs 10 and 20. 
The learning rate warms up for 6 epochs and decays from 0.8 to 8e-4.
In the large-margin fine-tuning stage, we fine-tune the model only on the TidyvoiceX-train set for 6 epochs, employing a margin of 0.5 and 6-second segments.

Across all stages, no speed perturbation is applied for data augmentation. 
Noise augmentation is applied with a probability of 0.6. 
An exponential decay scheduler is used for learning rate adjustment. The optimizer is AdamW for PTM-based SV models and SGD with Nesterov momentum, $m = 0.9$, for RedimNet. 
Weight decay is set to 2e-5 for all parameters of the PTM-based model. For RedimNet, the classification head uses a weight decay of 1e-4, while other components use 2e-5.

\section{Experiments} 

\subsection{Results} 

\begin{table*}[!t]
    \caption{
        Evaluation results of developed models on TidyVoiceX\_Dev, tv26\_eval-A and tv26\_eval-U, reported as EER(\%) / minDCF.
        Channel Concatenation is used for multi-scale feature aggregation.
        GMACs is measured on 2-second segments using the \texttt{THOP} library.
    }
    \vspace{-10pt} 
    \begin{center}
    \setlength{\tabcolsep}{6pt} 
    \setlength{\extrarowheight}{1pt} 
    \begin{tabular}{l|c|c|c|c|c}
    \toprule
     \multirow{2}{*}{\textbf{System}} &
    \multirow{2}{*}{\textbf{Params}} &
    \multirow{2}{*}{\textbf{GMACs}} &
    \textbf{TidyVoiceX\_Dev} &
    \textbf{tv26\_eval-A} &
    \textbf{tv26\_eval-U} \\
    & & & \textbf{EER(\%)} / \textbf{minDCF} & \textbf{EER(\%)} / \textbf{minDCF} & \textbf{EER(\%)} / \textbf{minDCF} \\

    \midrule
    
    \textbf{ReDimNet-B5}$_{SF2-C}$  & \multirow{2}{*}{9.2M} & \multirow{2}{*}{9.87} & 1.1353 / 0.6024 & - & - \\
    \hspace{0.75cm}+QMF  &  &  & - & 3.4699 / 0.2562 & 4.8389 / 0.2944 \\
    
    \textbf{ReDimNet-B6}$_{SF2-C}$  & \multirow{2}{*}{15M} & \multirow{2}{*}{20.27} & 0.9791 / 0.5887 & - & - \\
    \hspace{0.75cm}+QMF  &  &  & - & 2.5163 / 0.1892 & 3.4344 / 0.2237 \\

    \midrule
    \textbf{w2v-BERT 2.0 based SV}$_{AAM}$ & \multirow{2}{*}{580+13.7M} & \multirow{2}{*}{58.43}  & 1.0255 / 0.6085  & 3.3382 / 0.2328 & 4.5895 / 0.2676  \\
    \hspace{0.75cm}+QMF  &  &  & - & 2.7320 / 0.2524 & \textbf{2.8389} / 0.2789 \\

    \midrule
    \textbf{Fusion (without calibration)}  & - & - & \textbf{0.8573} / \textbf{0.5791} & - & - \\   
    \textbf{Fusion (with QMFs)}  & - & - & - & \textbf{2.2097} / \textbf{0.1797} & 2.9909 / \textbf{0.2045} \\     
    \bottomrule
    \end{tabular}
    \end{center}
    \label{tab:final}
\end{table*}

The performance of models is assessed on the tv26\_eval-A and tv26\_eval-U sets using the Equal Error Rate (EER) as the primary evaluation metric, and the minimum Detection Cost Function (minDCF) with $P_\text{target}= 0.01$ and $C_\text{FA}= C_\text{Miss}= 1$ as a secondary metric. 
We also report results on the TidyVoiceX\_Dev set for completeness and comparative analysis.
Table~\ref{tab:final} summarizes the performance of the proposed systems.
The w2v-BERT~2.0-based SV system trained with AAM loss achieves 1.03\% on the dev set, 3.34\% on eval-A, and 4.59\% on eval-U.
After applying the Quality Measure Functions (QMF)~\cite{li2023dku} back-end from the WeSpeaker~\cite{wang2024advancing} toolkit for score calibration, using TidyVoiceX\_Dev as the cohort, the EER on eval-A and eval-U is reduced to 2.73\% and 2.84\%.
For the lightweight ReDimNet models trained with SF2-C loss, ReDimNet-B5 and ReDimNet-B6 obtain 1.14\% and 0.98\% on the dev set, respectively.
With QMF, the ReDimNet-B5 system achieved EERs of 3.47\% on eval-A and 4.84\% on eval-U. The ReDimNet-B6 system achieved 2.52\% on eval-A and 3.43\% on eval-U.
Although the w2v-BERT~2.0-based SV system and the ReDimNet systems are trained on the same datasets, the former achieves a lower EER on eval-U, which involves unseen languages, indicating that the PTM-based SV system offers stronger generalization ability.

Finally, we perform simple score-level fusion by averaging the calibrated scores from all QMF-enhanced systems.
The fused system achieves the best overall performance among our submitted systems on the official evaluation sets, with an EER of 2.21\% on eval-A and 2.99\% on eval-U.

\subsection{Ablation studies}

\begin{table}[ht]
    \centering
    \caption{Performance comparison of different pre-trained models for speaker verification on the TidyVoiceX\_Dev set without score calibration, including their parameter sizes and the number of languages in their pre-training corpora.}
    \setlength{\tabcolsep}{1.5pt} 
    \setlength{\extrarowheight}{1pt} 
    \begin{tabular}{l|c|c|c}
    \toprule

    \multirow{2}{*}{\textbf{PTM}} &
    \multirow{2}{*}{\textbf{Params}} &
    \multirow{2}{*}{\textbf{\#Langs}} &
    \textbf{TidyVoiceX\_Dev} \\
    & & & \textbf{EER(\%)} / \textbf{minDCF}  \\
    
    \midrule
    Whisper Large-v3~\cite{radford2023robust}  & 635M & 99 & 3.1468 / 0.7352 \\
    
    \midrule
    XLS-R-300m~\cite{babu2021xls}              & 317M & 128 & 2.9394 / 0.6481  \\
    XLS-R-1b~\cite{babu2021xls}                & 965M & 128 & 3.0987 / \textbf{0.5663}  \\
    XLS-R-2b~\cite{babu2021xls}                & 2.16B & 128 & 3.1022 / 0.6598  \\
    \midrule
    MMS-300m~\cite{pratap2024scaling}          & 317M & 1406 & 3.2738 / 0.6676  \\
    MMS-1b~\cite{pratap2024scaling}            & 965M & 1406 & 3.4591 / 0.7124  \\
    
    \midrule
    \textbf{w2v-BERT 2.0}~\cite{barrault2023seamlessm4t} & 580M & 143 & \textbf{2.0927} / 0.7063  \\
    \bottomrule
    \end{tabular}
    \label{tab:abl_diff_ssl}
\end{table}

We experiment with several self-supervised pre-trained models of varying parameter scales, all trained on large-scale multilingual unlabeled speech, including the XLS-R series~\cite{babu2021xls}, the MMS series~\cite{pratap2024scaling}, and w2v-BERT 2.0~\cite{barrault2023seamlessm4t}, employing them as front-end feature extractors for speaker verification.
In addition, we evaluate a weakly supervised model, Whisper Large-v3~\cite{radford2023robust}, which is pre-trained on large-scale multilingual speech–text pairs.
Specifically, we utilize its encoder as the feature extraction front-end.
For multi-scale feature aggregation, we employ Channel Concatenation.
The results are summarized in Table~\ref{tab:abl_diff_ssl}. 
All SV models are fine-tuned using only the TidyVoiceX\_Train dataset and evaluated on the TidyVoiceX\_Dev set. 
Among the models tested, the w2v-BERT 2.0~\cite{barrault2023seamlessm4t} based SV system achieves the best performance, with an EER of 2.09\% on the dev set. 
The w2v-BERT 2.0 contains 580M parameters and is pre-trained using self-supervised objectives on unlabeled speech data covering 143 languages.

\begin{table}[ht]
    \centering
    \caption{Ablation study on multi-scale feature aggregation for w2v-BERT 2.0 based speaker verification models.}
    \setlength{\tabcolsep}{5.0pt} 
    \setlength{\extrarowheight}{1pt} 
    \begin{tabular}{l|c}
    \toprule
    \multirow{2}{*}{\textbf{Aggregation Method}} &
    \textbf{TidyVoiceX\_Dev} \\
    & \textbf{EER(\%)} / \textbf{minDCF}  \\
    \midrule
    Mean Fusion & 2.2134 / 0.7129  \\
    Layer-wise Weighted Fusion & 2.1856 / 0.7095 \\
    Channel-wise Weighted Fusion & 2.1771 / 0.7080 \\
    Channel Concatenation & \textbf{2.0927} / 0.7063 \\  
    Temporal Concatenation & 2.3210 / 0.7184 \\  
    Hierarchical Cross-Attention Fusion & 2.1543 / \textbf{0.6478} \\
    \bottomrule
    \end{tabular}
    \label{tab:abl_ssl_msfa}
\end{table}

We experiment with several multi-scale feature aggregation strategies as described in Section~\ref{sec:ssl_ptm_sv} on the w2v-BERT~2.0-based SV model, using only TidyVoiceX\_Train for fine-tuning. 
The results on TidyVoiceX\_Dev are reported in Table~\ref{tab:abl_ssl_msfa}. 
The performance differences among the methods are relatively small. Among them, Channel Concatenation achieves the lowest EER, while Hierarchical Cross-Attention Fusion attains the best minDCF.
Notably, Channel Concatenation introduces no additional trainable parameters.
We ultimately adopted Channel Concatenation for our final submission due to its superior EER performance and minimal architectural overhead.

\begin{table}[ht]
    \centering
    \caption{Ablation study of the w2v-BERT 2.0 based speaker verification models on TidyVoiceX\_Dev without score calibration.}
    \setlength{\tabcolsep}{4.5pt} 
    \setlength{\extrarowheight}{1pt} 
    \begin{tabular}{l|c|c}
    \toprule
    \multirow{2}{*}{\textbf{Training Stage}} &
    \multirow{2}{*}{\textbf{Training Data}} &
    \textbf{TidyVoiceX\_Dev} \\
    & & \textbf{EER(\%)} / \textbf{minDCF}  \\
    \midrule
    Pre-training & Full dataset  & 1.5411 / 0.6636 \\
    + Fine-tuning & Full dataset & 1.1380 / 0.6250 \\
    ++ LM-FT & TidyVoiceX\_Train & \textbf{1.0255} / \textbf{0.6085} \\  
    \bottomrule
    \end{tabular}
    \label{tab:abl_ssl_train}
\end{table}

To evaluate the contribution of each stage in the three-stage training paradigm described in Section~\ref{sec:train_paradigm}, we conduct an ablation study on the w2v-BERT~2.0-based speaker verification models.
The results are summarized in Table~\ref{tab:abl_ssl_train}, which compares three configurations incrementally incorporating each stage: using only the pre-trained model as a fixed front-end, then fine-tuning the entire model, and finally applying in-domain LM-FT.
The results confirm that each stage brings a performance gain.
In the first stage, the pre-trained w2v-BERT 2.0 is used as a frozen feature extractor while only the downstream modules are trained, achieving an EER of 1.54\% and a minDCF of 0.66.
Fine-tuning all model parameters in the second stage substantially improves performance, reducing the EER to 1.14\% and the minDCF to 0.63.
Finally, applying LM-FT on the target-domain TidyVoiceX\_Train data further refines the model, attaining the best performance with an EER of 1.03\% and a minDCF of 0.61. 
This progressive training strategy effectively adapts the SSL pre-trained models to speaker verification.

\section{Conclusions}

In this work, we designed a speaker verification framework based on self-supervised pre-trained models. 
The core idea is to leverage the robust acoustic representations learned from vast amounts of unlabeled multilingual audio through self-supervised learning, with the goal of building more generalizable and language-agnostic speaker recognition systems. 
We employed the w2v-BERT 2.0 as the pre-trained backbone and conducted a comparative analysis of various multi-scale feature aggregation strategies. 
Furthermore, we designed and implemented a three-stage training paradigm to progressively adapt the model. 
For comparison, we also trained lightweight ReDimNet models. 
The results indicate that, given the same training data, the w2v-BERT~2.0-based SV system demonstrates superior performance on evaluation sets containing unseen languages. 
Our final system fusion strategy achieved an EER of 2.21\% on the tv26\_eval-A set and 2.99\% on the tv26\_eval-U set on the TidyVoice2026 benchmark.

\clearpage 

\section{Generative AI Use Disclosure}

Generative AI assistants were employed solely for language editing and polishing of the manuscript. 
All scientific content, including the research concept, experimental design, model architecture, and analysis, was developed by the authors.

\bibliographystyle{IEEEtran}
\bibliography{mybib}

\end{document}